\let\emph\textit
\documentclass[conference,a4paper]{IEEEtran}
\usepackage{amsmath}
\usepackage{amsmath,bm}
\usepackage{amssymb}
\usepackage{multicol}
\usepackage{graphicx}
\usepackage{epstopdf}
\usepackage{booktabs}
\usepackage{color}
\usepackage{psfrag}
\usepackage{cleveref}
\usepackage{cite}
\usepackage{acronym}
\usepackage{arydshln}
\usepackage{stfloats}
\usepackage{subfigure}
\usepackage{caption}
\usepackage{setspace}
\usepackage{amssymb}
\usepackage{latexsym}

\DeclareMathAlphabet{\mathsfbr}{OT1}{cmss}{m}{n}
\SetMathAlphabet{\mathsfbr}{bold}{OT1}{cmss}{bx}{n}
\DeclareRobustCommand{\msf}[1]{%
	\ifcat\noexpand#1\relax\msfgreek{#1}\else\mathsfbr{#1}\fi
}

\makeatletter
\newcommand{\msfgreek}[1]{\csname s\expandafter\@gobble\string#1\endcsname}
\makeatother

\newenvironment{sproof}{{\indent \indent \it Sketch of Proof:\quad}}{\hfill $\square$\par}
\DeclareFontEncoding{LGR}{}{} 
\DeclareSymbolFont{sfgreek}{LGR}{cmss}{m}{n}
\SetSymbolFont{sfgreek}{bold}{LGR}{cmss}{bx}{n}
\DeclareMathSymbol{\salpha}{\mathord}{sfgreek}{`a}
\DeclareMathSymbol{\sbeta}{\mathord}{sfgreek}{`b}
\DeclareMathSymbol{\sgamma}{\mathord}{sfgreek}{`g}
\DeclareMathSymbol{\sdelta}{\mathord}{sfgreek}{`d}
\DeclareMathSymbol{\sepsilon}{\mathord}{sfgreek}{`e}
\DeclareMathSymbol{\szeta}{\mathord}{sfgreek}{`z}
\DeclareMathSymbol{\seta}{\mathord}{sfgreek}{`h}
\DeclareMathSymbol{\stheta}{\mathord}{sfgreek}{`j}
\DeclareMathSymbol{\siota}{\mathord}{sfgreek}{`i}
\DeclareMathSymbol{\skappa}{\mathord}{sfgreek}{`k}
\DeclareMathSymbol{\slambda}{\mathord}{sfgreek}{`l}
\DeclareMathSymbol{\smu}{\mathord}{sfgreek}{`m}
\DeclareMathSymbol{\snu}{\mathord}{sfgreek}{`n}
\DeclareMathSymbol{\sxi}{\mathord}{sfgreek}{`x}
\DeclareMathSymbol{\somicron}{\mathord}{sfgreek}{`o}
\DeclareMathSymbol{\spi}{\mathord}{sfgreek}{`p}
\DeclareMathSymbol{\srho}{\mathord}{sfgreek}{`r}
\DeclareMathSymbol{\ssigma}{\mathord}{sfgreek}{`s}
\DeclareMathSymbol{\stau}{\mathord}{sfgreek}{`t}
\DeclareMathSymbol{\supsilon}{\mathord}{sfgreek}{`u}
\DeclareMathSymbol{\sphi}{\mathord}{sfgreek}{`f}
\DeclareMathSymbol{\schi}{\mathord}{sfgreek}{`q}
\DeclareMathSymbol{\spsi}{\mathord}{sfgreek}{`y}
\DeclareMathSymbol{\somega}{\mathord}{sfgreek}{`w}

\DeclareMathSymbol{\svarsigma}{\mathord}{sfgreek}{`c}

\DeclareMathSymbol{\sGamma}{\mathalpha}{sfgreek}{`G}
\DeclareMathSymbol{\sDelta}{\mathalpha}{sfgreek}{`D}
\DeclareMathSymbol{\sTheta}{\mathalpha}{sfgreek}{`J}
\DeclareMathSymbol{\sLambda}{\mathalpha}{sfgreek}{`L}
\DeclareMathSymbol{\sXi}{\mathalpha}{sfgreek}{`X}
\DeclareMathSymbol{\sPi}{\mathalpha}{sfgreek}{`P}
\DeclareMathSymbol{\sSigma}{\mathalpha}{sfgreek}{`S}
\DeclareMathSymbol{\sUpsilon}{\mathalpha}{sfgreek}{`U}
\DeclareMathSymbol{\sPhi}{\mathalpha}{sfgreek}{`F}
\DeclareMathSymbol{\sPsi}{\mathalpha}{sfgreek}{`Y}
\DeclareMathSymbol{\sOmega}{\mathalpha}{sfgreek}{`W}

\DeclareRobustCommand{\mcal}[1]{%
	\ifcat\noexpand#1\relax\mathnormal{#1}\else\cal{#1}\fi
}
\DeclareRobustCommand{\BM}[1]{%
	\ifcat\noexpand#1\relax\bm{\boldUppercaseItalicGreek{#1}}\else\bm{#1}\fi
}
\makeatletter
\newcommand{\boldUppercaseItalicGreek}[1]{\csname var\expandafter\@gobble\string#1\endcsname}
\makeatother
\newcommand{\RS}[1]{\MakeUppercase{\msf{#1}}} 

\newcommand{\V}[1]{\bm{#1}} 

\newtheorem{proposition}{Proposition}
\newtheorem{remark}{Remark}
\newtheorem{definition}{Definition}

\ifCLASSINFOpdf
\else
\fi
\begin{document}

%
\title{On the Performance Tradeoff of an ISAC System with Finite Blocklength}

\author{\IEEEauthorblockN{Xiao Shen*, $\text{Na\ Zhao}^\dagger$, and Yuan Shen*}
\IEEEauthorblockA{* Department of Electronic Engineering, Tsinghua University, Beijing 100084, China\\Tsinghua National Laboratory for Information Science and Technology\\
$^\dagger$ School of Electronic and Information Engineering, Beihang University, Beijing, 100191, China\\
Email: shenx20@mails.tsinghua.edu.cn, na\_zhao@buaa.edu.cn, shenyuan\_ee@tsinghua.edu.cn}

}


%


\maketitle

\begin{abstract}
Integrated sensing and communication (ISAC) has been proposed as a promising paradigm in the future wireless networks, where the spectral and hardware resources are shared to provide a considerable performance gain. It is essential to understand how sensing and communication (S\&C) influences each other to guide the practical algorithm and system design in ISAC. In this paper, we investigate the performance tradeoff between S\&C in a single-input single-output (SISO) ISAC system with finite blocklength. In particular, we present the system model and the ISAC scheme, after which the rate-error tradeoff is introduced as the performance metric. Then we derive the achievability and converse bounds for the rate-error tradeoff, determining the boundary of the joint S\&C performance. Furthermore, we develop the asymptotic analysis at large blocklength regime, where the performance tradeoff between  S\&C is proved to vanish as the blocklength tends to infinity. Finally, our theoretical analysis is consolidated by simulation results. 
\end{abstract}



%
\IEEEpeerreviewmaketitle

\section{Introduction}

Recent years have witnessed the rapid development of 5G technologies in modern civil and military applications such as intelligent vehicular networks and rescue operations, which leads to the severe congestion of the spectral and hardware resources\cite{HugKawSim:J15}. To improve the resource utilization and reduce costs, ISAC is emerging as the key technique in the next-generation wireless networks\cite{LiuCuiMas:J22,StuWie:J11}. Compared with the traditional wireless networks where the communication and sensing systems are designed separately, ISAC applies the 
integrated signals to perform the dual-functions simultaneously, where the performance gain comes from the sharing of spectrum and hardwares. 

There exist many crucial problems in the ISAC research including the theoretical framework, the system protocol and the signal processing algorithms. The most important and challenging one among them is the characterization of the performance tradeoff between S\&C, which comes from the signal sharing in ISAC\cite{ZhaRahWu:J21}. From the perspective of communication, we aim to add more randomness to the transmitted signals to carry more information, while the deterministic waveforms are more conducive to improving the estimation accuracy of our interested parameters from the perspective of sensing. The performance tradeoff analysis can not only contribute to determining the fundamental limits of the dual-functions, but also guide the design and operation of practical ISAC systems. Therefore, extensive works have been dedicated to address this issue.

In \cite{AhmKobWig:J22}, the authors characterize the capacity-distortion-cost tradeoff in the ISAC systems where sensing refers to the state estimation based on the state-dependent channel feedbacks, and a modified Blahut-Arimoto algorithm is proposed to numerically depict the tradeoff region. In \cite{ChaMoeHim:J17}, the performance tradeoff between the radar receiver and the communication receiver is investigated in a ISAC system, where the boundaries of probability-rate regions are derived. Recently, the authors in \cite{XioLiuCui:J22} has determined the S\&C performance at the two corner points of the CRB-rate region to reveal a two-fold tradeoff in ISAC systems.

Although progress has been made in terms of establishing the theoretical foundation of ISAC, there exists a common limitation in the above studies. Most existing works use Shannon channel capacity as the performance metric for the communication rate, which is an asymptotic result with the code blocklength approaching infinity. However, the evaluation of sensing performance is meaningless with infinite blocklength since the estimation error tends zero with infinite signal energy. Therefore, it is essential to analyze the performance tradeoff between S\&C in the finite blocklength regime, which motivates our work.

In this paper, we characterize the performance tradeoff between S\&C in a SISO ISAC system with finite blocklength. First in Section II, we present the system model including the ISAC scheme and performance metrics, where the rate-error region is defined to evaluate the tradeoff. Then in Section III, we derive the achievability and converse bounds for the rate-error tradeoff, after which the asymptotic analysis is performed to show that the tradeoff vanishes as the blocklength increases. In section IV, the theoretical results are verified by the numerical experiments. Finally, Section V concludes this paper.

\section{System Model}
In this section, we first present our signal model and ISAC scheme, after which the performance metrics for communication and sensing are introduced to characterize the rate-error tradeoff.

Consider a SISO ISAC signal model given by
\begin{equation}
\label{Channel}
    \mathbf{y} = h\mathbf{x}+\mathbf{n}
\end{equation}
where $\mathbf{x}\in \mathbb{C}^N$ is the transmitted random communication symbol and $N$ denotes the blocklength. The notation $\mathbb{C}^N$ denotes $N$-dimensional complex Euclidean space where the superscript is removed with $N=1$, while $\mathbb{R}^N$ refers to the $N$-dimensional real Euclidean space. We denote by $\mathbf{y}\in \mathbb{C}^N$ the received signal and $\mathbf{n}\in \mathbb{C}^N$ the circularly symmetric complex Gaussian noise with zero mean, i.e., $\mathbf{n}\sim\mathcal{CN}(0,\sigma^2\V{I}_N)$. The scalar $h \in \mathbb{C}$ is the unknown but deterministic channel coefficient determined by the sensing parameters, which is assumed to be constant in all $N$ channel uses since the sensing parameters such as target positions and velocities remain stable during the communication process in most ISAC systems. The goal of ISAC is to simultaneously recover the communication message and estimate the sensing parameters based on the dual-functional signal. 
\subsection{ISAC Scheme} 
\begin{figure}[t]
    \centering
    \includegraphics[width = 8.5cm, height = 3cm]{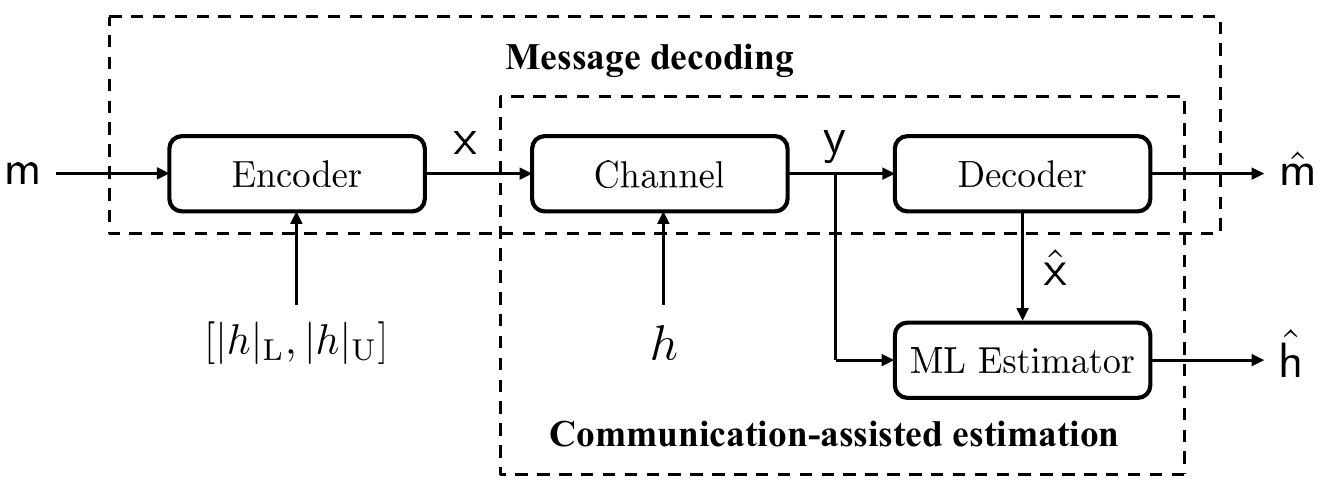}
    \caption{ The block diagram of our proposed ISAC scheme, which consists of the message-decoding step and the communication-assisted-estimation step.}
    \label{Fig_scheme}
\end{figure}

To analyze the performance trade-off between S\&C, we first present the mathematical formulation for our ISAC system, which is based on a two-step scheme including message decoding and communication-assisted estimation shown in Fig. \ref{Fig_scheme}.

The message-decoding step aims to recover the transmitted communication symbol based on the received signal $\mathbf{y}(\mathbf{x},h)$, which determines the communication performance of our ISAC system. In particular, we applies the $(N,M,\epsilon)$ code introduced in \cite{PolPooVer:J10}, which includes
\begin{enumerate}
    \item A message set $\mathcal{M} = \{1,2,\ldots,M\}$ with equiprobable messages;
    \item An encoder which maps the message $\mathrm{m}\in\mathcal{M}$ to the codewords $\V{x}_m \in \mathcal{X} = \{\V{x}_1,\V{x}_2,\ldots,\V{x}_M\}$. We assume that the channel gain is confined to a certain set, i.e., $|h| \in [|h|_\mathrm{L},|h|_\mathrm{U}]$, which is known to the encoder\footnote{In communication theory, the channel coefficient is usually modeled as a random variable with certain prior distribution, while it is mainly determined by the deterministic but unknown sensing parameters in ISAC systems. Therefore, the encoder is assumed to have knowledge of the uncertainty set of the channel gain to design the codebook in the ISAC settings, which is also feasible in practical systems}. The notation $|h|$ denotes the absolute value of the complex number $h$. Therefore, the encoder can be expressed by
    \begin{equation}
        f: \mathcal{M}\times\mathbb{R}^2\mapsto \mathcal{X},\ m\times [|h|_\mathrm{L},|h|_\mathrm{U}]\to\V{x}_m.
    \end{equation}
    Furthermore, the codewords satisfy the power constraint
    \begin{equation}
    \label{Power-constraint}
        \|\V{x}_i\|_2^2 \leq N\rho, \quad i=1,\ldots,M
    \end{equation}
    where the constant $\rho$ is the per-codeword power budget;
    \item A decoder which maps the received signal to the message, i.e.,
    \begin{equation}
        g: \mathbb{C}^N \mapsto \mathcal{M}, \ \mathbf{y}\to m.
    \end{equation}
    Furthermore, the decoder satisfies
    \begin{equation}
        \mathbb{P}\{g(\mathbf{y})\neq \mathrm{m}\} \leq \epsilon.
    \end{equation}
    where $\mathbb{P}(\RS{X})$ denotes the probability of the random set $\RS{X}$.
\end{enumerate}

The communication-assisted-estimation step aims to estimate the sensing parameters based on the received signal, which determines the sensing performance of our ISAC system. For simplicity of analysis, we focus on the estimation of the channel coefficient itself in this paper, while the analysis of estimating general sensing parameters is left for our future work. In particular, we first reconstruct the communication symbol as $\hat{\mathbf{x}} = g(\mathbf{y})$. Then we apply a maximum-likelihood (ML) estimator to estimate the channel coefficient since the ML estimator can asymptotically achieve the Cram\'er-Rao lower bound\cite{VanBel:B68}. After some algebra, we can obtain that
\begin{equation}
\label{ML-estimator}
     \hat{\mathrm{h}} = \hat{\mathrm{h}}_\mathrm{ML}(\hat{\mathbf{x}},\mathbf{y}) = \frac{\hat{\mathbf{x}}^\mathrm{H}\mathbf{y}}{\|\hat{\mathbf{x}}\|_2^2}.
\end{equation}
where the notation $\V{x}^\mathrm{H}$ denotes the Hermitian transposition of complex vector $\V{x}$. This ISAC scheme takes full advantage of the communication result to improve the sensing performance of the channel coefficient, which is widely applied in the practical ISAC systems\cite{ZhaLiuMas:J21}.
\subsection{Performance Metric}
In this subsection, we give a brief introduction on the traditional performance metrics of communication and sensing, after which the rate-error region is introduced to characterize the performance tradeoff. 

The performance metric for communication systems with the quasi-static channel is defined as the achievable communication rate for the $(N,M,\epsilon)$ code, i.e.,
\begin{equation}
    R = \frac{\log_2 M}{N}: \exists(N, M, \epsilon) \text { code. }
\end{equation}

The performance metric for sensing systems is characterized by the mean squared error (MSE) of the ML estimator, i.e., given any $(N, M, \epsilon)$ code, 
\begin{equation}
\label{MSE}
    e = \mathbb{E}_{\mathbf{y},\mathbf{x}}\bigg\{|h-\hat{\mathrm{h}}_\mathrm{ML}(\hat{\mathbf{x}},\mathbf{y})|^2\bigg\}.
\end{equation}
where the notation $\mathbb{E}_{\mathbf{x}}\left\{\cdot\right\}$ refers to the expectation with respect to the random variable $\mathbf{x}$.

When both the communication and sensing performance are taken into consideration, we define the rate-error region as
\begin{equation}
\mathcal{F}(N,\epsilon) = \bigg\{(R,e):\exists(N, M, \epsilon) \text { code}\bigg\}
\end{equation}
which collects all the feasible pairs of the communication rate and sensing error achieved by the $(N,M,\epsilon)$ code. Then the boundary of $\mathcal{F}(N,\epsilon)$ reveals the optimal performance of communication or sensing when the other one meets certain minimum requirements, which characterizes the performance tradeoff. Therefore, we focus on determining the boundary of the rate-error region in the following sections of this paper. Namely, we aim to obtain
\begin{equation}
    R^\star(N,\epsilon,D) = \sup\bigg\{R:(R,e)\in \mathcal{F}(N,\epsilon), e\leq D\bigg\}
\end{equation}
where $D$ denotes the minimum requirements for the sensing performance, and $R^\star$ denotes the rate-error tradeoff. When $D$ tends to infinity, the rate-error tradeoff approaches the maximal achievable rate of the quasi-static channel regardless of sensing performance, i.e., 
\begin{equation}
\lim_{D\to\infty} R^\star(N,\epsilon,D) = R_\mathrm{com}^\star(N,\epsilon)
\end{equation}
which has been widely investigated in the communication theory\cite{PolPooVer:J10,YanDurKoc:C13,YanDurKoc:J10}.

\begin{remark}
    In most existing researches on the performance tradeoff between S\&C, the influence of the blocklength $N$ and the probability of decoding error $\epsilon$ are not taken into consideration since the estimation error is independent of the decoder outputs where communication and sensing are separately performed at the receiver and the transmitter, respectively\cite{AhmKobWig:J22,ChaMoeHim:J17,XioLiuCui:J22}. However, when dual functionalities are required at the receiver concurrently, such as in the intelligent vehicular networks and other cooperative applications, existing tradeoff analysis is inapplicable. As will be shown later, the blocklength and the probability of decoding error induce a tighter connection for S\&C and require in-depth rate-error analysis.
\end{remark}

\section{Rate-error Tradeoff Analysis}
In this section we characterize the rate-error tradeoff in our ISAC systems. Note that even the exact expression of the maximal achievable rate regardless of sensing performance is intractable with a fixed blocklength and the probability of decoding error. We derive the achievability (lower) and converse (upper) bound of $R^\star$ to characterize the performance tradeoff between S\&C, and discuss the asymptotic property of these two bounds when the blocklength $N$ tends to infinity.

\subsection{Achievability Bound}
In this subsection, we present the achievability bound of the rate-error tradeoff $R^\star(N,\epsilon,D)$. 

First, we tighten the inequality power constraint \eqref{Power-constraint} into the equality one, i.e.,  $\|\V{x}_i\|_2^2 = N\rho, \ i=1,\ldots,M$. We denote by $\tilde{R}^\star(N,\epsilon,D)$ the rate-error tradeoff with the tightened power constraint. Then $\tilde{R}^\star(N,\epsilon,D)$ must be a lower bound of $R^\star(N,\epsilon,D)$. Furthermore, when we consider the maximal communication rate regardless of sensing performance, it is proved that\cite{PolPooVer:J10}
\begin{equation}
    \tilde{R}_\mathrm{com}^\star(N,\epsilon) \leq R_\mathrm{com}^\star(N,\epsilon)\leq \frac{N+1}{N}\tilde{R}_\mathrm{com}^\star(N+1,\epsilon).
\end{equation}
which implies that the approximation is asymptotically tight. Therefore, we only need to determine the achievability bound for $\tilde{R}^\star(N,\epsilon,D)$, where the codewords are distributed on the complex hypersphere $\mathcal{S}^N = \{\V{x}\in\mathbb{C}^N:\|\V{x}\|^2 = N\rho\}$.

Then we investigate the relationship between the MSE given by \eqref{MSE} and the feasible set of the codewords. In particular, given a set $\mathcal{W}\subseteq \mathcal{S}^N$, we introduce the definition of the maximal bias as follows.
\begin{definition}
    The maximal bias with respect to any subset $\mathcal{W}\subseteq \mathcal{S}_N$ is defined as
    \begin{equation}
    \Delta_{\mathcal{W}} = \sup \bigg\{\bigg|1-\frac{\V{u}^\mathrm{H}\V{v}}{N\rho}\bigg|: \V{u},\V{v}\in \mathcal{W}\bigg\}.
    \end{equation}
\end{definition}

\begin{remark}
Recalling the expression shown in \eqref{ML-estimator}, we find that $\Delta_{\mathcal{W}}$ equals to the maximum mean error of any ML estimator with both the transmitted symbol $\mathbf{x}$ and the reconstructed symbol $\hat{\mathbf{x}}$ distributed in the set $\mathcal{W}$. Therefore, the maximal bias characterizes the diversity of the codewords within a certain set on the hypersphere with respect to channel estimation, which attains the maximum value $\Delta_{\mathcal{W}} = 2$ with $\mathcal{W} = \mathcal{S}^N$ and the minimum value $\Delta_{\mathcal{W}} = 0$ with $\mathcal{W}$ as any single point set.
\end{remark}

Then we can derive an upper bound of the MSE when the codewords are distributed in the set $\mathcal{W}$, which is formulated as the following proposition.
\begin{proposition}
\label{Pro:Error-upperbound}
For the $(N,M,\epsilon)$ code where the codewords are distributed in the set $\mathcal{W}\subseteq \mathcal{S}^N$, the MSE for the ML estimator is upper-bounded by
\begin{equation}
\mathrm{MSE}\leq \frac{\sigma^2}{N\rho}+\epsilon|h|_\mathrm{U}^2\Delta_{\mathcal{W}}^2+\frac{2\sigma\sqrt{\epsilon}|h|_\mathrm{U}}{\sqrt{N\rho}}\Delta_{\mathcal{W}}
\end{equation}
\end{proposition}

\begin{remark}
According to proposition \ref{Pro:Error-upperbound}, the upper bound of MSE can be divided into two parts: the first part refers to the sensing performance in the ideal case where the reconstructed symbol equals to the transmitted communication symbol. The second part refers to the sensing performance with decoding error, which consists of the bias term and the cross term since $\hat{\mathrm{h}}_\mathrm{ML}$ is not an unbiased estimator of $h$ with $\hat{\mathbf{x}}\neq \mathbf{x}$. In the radar-based ISAC systems where the transmitted symbol is known to the estimator, i.e., $\epsilon = 0$, we obtain that $\mathrm{MSE} = \sigma^2/N\rho$ which coincides with the existing theoretical results. 
\end{remark}

With Proposition \ref{Pro:Error-upperbound}, we find that the the MSE is also controlled by the maximal bias of the codeword set, which provides useful insights for the following analysis. Then we derive the achievability bound for the rate-error tradeoff with the tightened power constraint. In particular, we first prove that $\tilde{R}^\star(N,\epsilon,D)$  saturates when $D$ exceeds certain threshold.
\begin{proposition}
\label{Pro:Saturate}
    The rate-error tradeoff $\tilde{R}^\star(N,\epsilon,D)$ satisfies
    \begin{equation}
         \tilde{R}^\star(N,\epsilon,D) = \tilde{R}^\star(N,\epsilon,D_\mathrm{m}) = \tilde{R}_\mathrm{com}^\star(N,\epsilon) ,\quad \forall D\geq D_\mathrm{m} 
    \end{equation}
    where $\tilde{R}_\mathrm{com}^\star(N,\epsilon)$ denotes the maximal achievable rate regardless of the sensing performance with the tightened power constraint, which has been investigated in \cite{YanDurKoc:C13,YanDurKoc:J10}. The threshold $D_\mathrm{m}$ is given by
    \begin{equation}
    \label{Maximal-error}
        D_\mathrm{m} = \frac{\sigma^2}{N\rho}+4\epsilon|h|_\mathrm{U}^2+\frac{4\sigma\sqrt{\epsilon}|h|_\mathrm{U}}{\sqrt{N\rho}}.
    \end{equation}
\end{proposition}

With Proposition \ref{Pro:Saturate}, we only need to derive the achievability bound in the case $D<D_\mathrm{m}$, which is shown in the following proposition
\begin{proposition}
\label{Pro:achievability-bound}
    For ${\sigma^2}/{N\rho}< D<D_\mathrm{m}$, the rate-error tradeoff $R^\star(N,\epsilon,D)$ is lower-bounded by 
    \begin{equation}
    \tilde{R}^\star(N,\epsilon,D) \geq \max\{\tilde{R}_\mathrm{com}^\star(N,\epsilon)+\frac{\log_2 \gamma_\mathrm{L}}{N},0\}
    \end{equation}
    where $\gamma_\mathrm{L} \in (0,1)$ is given by
    \begin{equation}
    \label{Achievability-rate-loss}
        \gamma_\mathrm{L} = 
        \frac{1}{2}\mathrm{I}_{\sin^2(\phi_\mathrm{L}/2)}(\frac{2N-1}{2},\frac{1}{2}).
    \end{equation}
    The function $\mathrm{I}_x(a,b)$ is the regularized incomplete beta function. The angle $\phi_\mathrm{L}\in[0,\pi)$ is determined by
    \begin{equation}
        \phi_\mathrm{L} = \arccos\bigg(\frac{2-\Delta_{\mathcal{W}_\mathrm{L}}^2}{2}\bigg)
    \end{equation}
    where the maximal bias $\Delta_{\mathcal{W}_\mathrm{L}}$ is given by
    \begin{equation}
        \Delta_{\mathcal{W}_\mathrm{L}} = \frac{\sqrt{DN\rho}-\sigma}{|h|_\mathrm{U}\sqrt{\epsilon N\rho}}.
    \end{equation}
    Specifically, there exists $\tilde{R}^\star(N,\epsilon,D) = 0$ for $0\leq D\leq\frac{\sigma^2}{N\rho}$.
\end{proposition}

\begin{sproof}
As for the analysis of the achievability bound, we guarantee the sensing performance through constraining the feasible set of the $(N,M,\epsilon)$ codewords, which is shown in Fig. 2(a). In particular, the codewords can take values on the entire hypersphere with $D\geq D_\mathrm{m}$, which implies that $R^\mathrm{L}(N,\epsilon,D) = \tilde{R}^\mathrm{L}_\mathrm{com}(N,\epsilon)$. When there exists $\sigma^2/N\rho\leq D <D_\mathrm{m}$, we restrict the codewords to the set $\mathcal{W}\subset\mathcal{S}^N$ with $\Delta_\mathcal{W} \leq \Delta_\mathcal{W}^\mathrm{L}$ to meet the minimum sensing performance requirement, where the largest feasible set $\mathcal{W}$ is a hyperspherical cap. The coefficient $\gamma_\mathrm{L}$ can be viewed as the area ratio of the hyperspherical cap to the hypersphere. Details are omitted due to the lack of space.
\end{sproof}

\begin{remark}
With Proposition \ref{Pro:Saturate} and \ref{Pro:achievability-bound}, we can obtain the achievability bound for the rate-error tradeoff $R^\star(N,\epsilon,D)$. We denote by $\tilde{R}^\mathrm{L}_\mathrm{com}(N,\epsilon)$ as the achievability bound of $\tilde{R}^\star_\mathrm{com}(N,\epsilon)$ determined by $|h|_\mathrm{L}$, the expression of which is provided in \cite{YanDurKoc:C13}. Then the achievability bound $R^\mathrm{L}(N,\epsilon,D)$ of $R^\star(N,\epsilon,D)$ is given by
\begin{equation}
    R^\mathrm{L} = \begin{cases}
    0 & 0\leq D\leq{\sigma^2}/{N\rho},\\
    \max\{\tilde{R}_\mathrm{com}^\mathrm{L}(N,\epsilon)+\frac{\log_2 \gamma_\mathrm{L}}{N},0\} & {\sigma^2}/{N\rho}< D <D_\mathrm{m},\\
    \tilde{R}^\mathrm{L}_\mathrm{com}(N,\epsilon) & D\geq D_\mathrm{m}.
    \end{cases}
\end{equation}
\end{remark}
\hspace{-1cm}

\begin{figure}[t]
    \centering
    \subfigbottomskip=2pt 
	\subfigcapskip=-5pt 
    	\subfigure[Achievability Bound]{
		\includegraphics[height = 3cm]{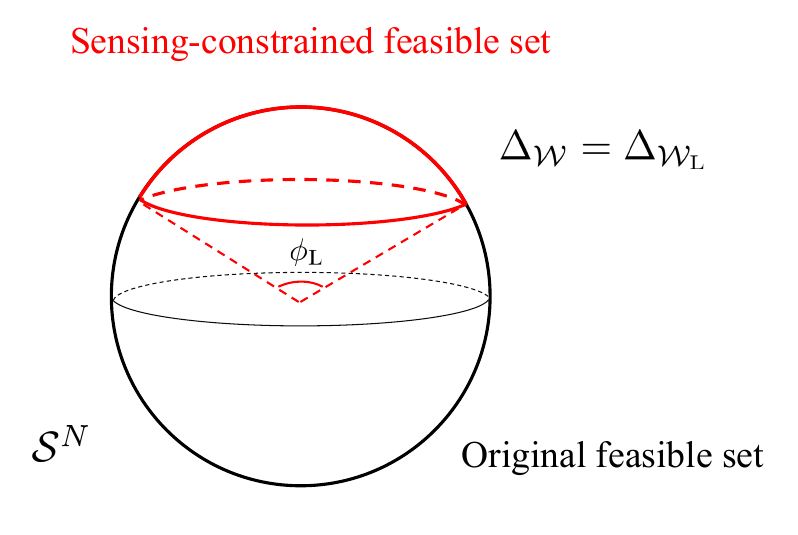}}
	\subfigure[Converse Bound]{
		\includegraphics[height = 3cm]{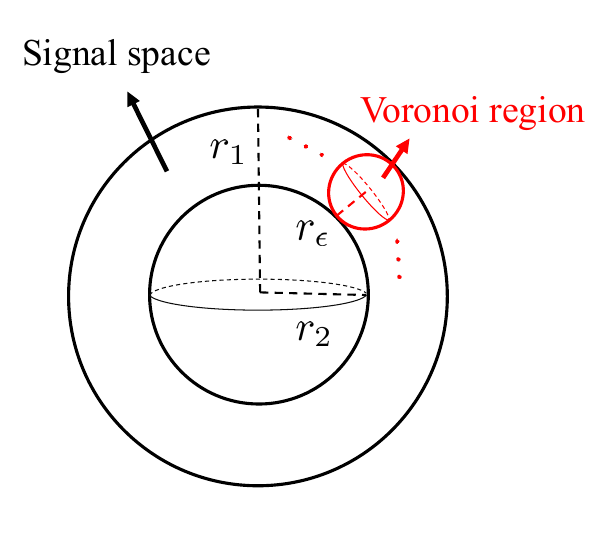}}
    \caption{ Geometric illustrations of the achievability and converse bounds. (a): For the achievability bound, the sensing performance $D$ constrains the original codeword space (black sphere) to the feasible set $\Delta_\mathcal{W}$ (red spherical cap). (b): For the converse bound, the maximal rate is obtained by packing the Voronoi region (red sphere of radius $r_\epsilon$) into the signal space, which is smaller than the black sphere of radius $r_1$ but bigger than the black spherical shell of radius $r_1$ and $r_2$.}
    \label{Fig_rate_error}
\end{figure}

\subsection{Converse Bound}
In this subsection, we present the converse bound of the rate-error tradeoff $R^\star(N,\epsilon,D)$. 


Note that the MSE of the ISAC system is lower-bounded by
\begin{equation}
    \mathrm{MSE}\geq \mathbb{E}_{\mathbf{x}}\bigg\{\frac{\sigma^2}{\|\mathbf{x}\|_2^2}\bigg\} \geq \frac{\sigma^2}{\mathbb{E}_{\mathbf{x}}\{\|\mathbf{x}\|_2^2\}}.
\end{equation}
Then for any $(R,e)\in \mathcal{F}(N,\epsilon)$, the $(N,M,\epsilon)$ code must at least satisfy the average power constraint given by
\begin{equation}
\label{Average-power-constraint}
\mathbb{E}_{\mathbf{x}}\{\|\mathbf{x}\|_2^2\} \geq \frac{\sigma^2}{D}.
\end{equation}

Therefore, we denote by $R^\mathrm{U}(N,\epsilon,D)$ the maximal achievable communication rate for the $(N,M,\epsilon)$ code satisfying power constraints \eqref{Power-constraint} and \eqref{Average-power-constraint}, which is a converse bound for the rate-error tradeoff $R^\star(N,\epsilon,D)$. Note that the expectation term in \eqref{Average-power-constraint} makes it difficult to obtain the exact expression of $R^\mathrm{U}(N,\epsilon,D)$. We provide a approximation of it in the following proposition.

\begin{proposition}
\label{Pro:Converse-bound}
The maximal achievable communication rate $R^\mathrm{U}(N,\epsilon,D)$ for the $(N,M,\epsilon)$ code satisfying the power constraints  \eqref{Power-constraint} and \eqref{Average-power-constraint} satisfies
\begin{equation}
\label{Converse-bound}
R^\mathrm{U}_\mathrm{com}(N,\epsilon)+\frac{\log_2(1-\gamma_\mathrm{U}^{2N}) }{N}  \leq R^\mathrm{U}(N,\epsilon,D) \leq R^\mathrm{U}_\mathrm{com}(N,\epsilon)
\end{equation}
where $\gamma_\mathrm{U}$ is given by $\gamma_\mathrm{U} = r_2/r_1$. The coefficient $r_1,r_2$ are given by
\begin{equation}
{r}_1 = \sqrt{|h|_\mathrm{U}^2N\rho+N\sigma^2},\ {r}_2 = \sqrt{\frac{|h|_\mathrm{L}^2\sigma^2}{D}+N\sigma^2},
\end{equation}
respectively. We denote $R^\mathrm{U}_\mathrm{com}(N,\epsilon)$ as the converse bound of the maximal communication rate regardless of sensing performance given by
\begin{equation}
    R^\mathrm{U}_\mathrm{com}(N,\epsilon) = \log_2\frac{r_1}{r_\epsilon}
\end{equation}
where the coefficient $r_\epsilon$ is the smallest $r$ such that the following inequality holds 
\begin{equation}
\label{r-epsilon}
    \mathbb{P}\{\mathbf{n}\notin \mathcal{B}_r^N\} \leq \epsilon
\end{equation}
where $\mathcal{B}_r^N\subset\mathbb{C}^N$ is the complex hyperball with radius $r$, i.e.,
$\mathcal{B}_r^N = \{\V{x}\in\mathbb{C}^N:\|\V{x}\|_2\leq r\}$.
\end{proposition}

\begin{sproof}
 According to the hypothesis testing theory, the optimal decoder with equiprobable messages is the maximum-likelihood decoder, the decoding regions of which are called the \emph{Voronoi regions}\cite{Sha:J59}. Then Proposition \ref{Pro:Converse-bound} is inspired from the idea of sphere packing where the Voronoi region is treated as the hyperball with radius $r_\epsilon$. As is shown in Fig. 2(b), the first inequality in \eqref{Converse-bound} is obtained by the sphere packing within the hyperspherical shell of radius $r_1$ and $r_2$, while the second one is obtained by the sphere packing within the entire hyperball of radius $r_1$. Details are omitted due to the lack of space.
\end{sproof}

\begin{remark}
 Proposition \ref{Pro:Converse-bound} provides a converse bound of the rate-error tradeoff $R^\star(N,\epsilon,D)$ which takes the influence of $D$ into consideration. However, $R^\mathrm{U}(N,\epsilon,D)$ is nearly independent of $D$ with large $N$ since the term $\log_2(1-\gamma_\mathrm{U}^{2N})$ decays exponentially with the blocklength, which equals to the converse bound of the maximal communication rate regard of the sensing performance given by the RHS of \eqref{Converse-bound}, i.e., $R^\mathrm{U}(N,\epsilon,D) \approx R^\mathrm{U}_\mathrm{com}(N,\epsilon) = 2\log_2(r_1/r_\epsilon)$. Therefore, the derivation of a tighter converse bound is still needed to give a more accurate characterization on the performance tradeoff between S\&C, which is a challenge left as our future work.
\end{remark}

\subsection{Asymptotic Analysis}
In this subsection, we present the asymptotic analysis for the achievability and converse bounds of the rate-error tradeoff $R^\star$ when the blocklength $N$ tends to infinity.

According to the theoretical analysis in \cite{YanDurKoc:C13}, the achievability bound $\tilde{R}^\mathrm{L}_\mathrm{com}(N,\epsilon)$ of the maximal communication rate regardless of sensing performance is proved to satisfy
\begin{equation}
    \lim_{N\to\infty}\tilde{R}^\mathrm{L}_\mathrm{com}(N,\epsilon) = \log_2(1+\frac{N\rho|h|_\mathrm{L}^2}{\sigma^2}) 
\end{equation}
for any $\epsilon\in(0,1/2)$. Note that there exists $\lim_{N\to\infty} D_\mathrm{m} = 4\epsilon|h|_\mathrm{U}^2$ according to \eqref{Maximal-error}. We can obtain
\begin{equation}
    \lim_{N\to \infty}R^\mathrm{L}(N,\epsilon,D) =  \lim_{N\to\infty}\tilde{R}^\mathrm{L}_\mathrm{com}(N,\epsilon) = \log_2(1+\frac{N\rho|h|_\mathrm{L}^2}{\sigma^2}) 
\end{equation}
for any $D> 4\epsilon|h|_\mathrm{U}^2$ and $\epsilon\in(0,1/2)$. 

Then we focus on the asymptotic analysis for the converse bound $R^\mathrm{U}(N,\epsilon,D)$. According to analysis to that in \cite{FozMcLSch:J03}, there exists
\begin{equation}
    \lim_{N\to \infty}\frac{r_\epsilon}{\sqrt{N\sigma^2}} = 1,\ \forall \epsilon \in (0,\frac{1}{2}).
\end{equation}
Therefore, the asymptotic expression of the converse bound $R^\mathrm{U}(N.\epsilon.D)$ is given by
\begin{equation}
    \lim_{N\to \infty}R^\mathrm{U}(N,\epsilon,D) = \lim_{N\to \infty}R^\mathrm{U}_\mathrm{com}(N,\epsilon) =\log_2(1+\frac{N\rho|h|_\mathrm{U}^2}{\sigma^2})
\end{equation}
for any $\epsilon\in(0,1/2)$.

According to the above theoretical analysis, we find that the performance tradeoff between  S\&C vanishes as the blocklength $N$ increases. We provide a simple interpretation for this phenomenon. Consider an ISAC system where the $(N,M,\epsilon)$ code consists of two parts: the first part of length $\sqrt{N}$ is fixed as the pilot data while the other part of length $N-\sqrt{N}$ is treated as the communication data. The sensing performance of this ISAC system should at least achieve $D_0 = \sigma^2/\sqrt{N}\rho$, which is the MSE of the channel coefficient $h$ obtained from the pilot data only. As $N$ tends to infinity, we have $D_0\to 0$ which implies that the sensing requirement can always be met by the pilot data with large blocklength $N$. Note that the pilot data of length $\sqrt{N}$ will not influence the maximal communication rate asymptotically since there exists $\lim_{N\to\infty} \sqrt{N}/N = 0$. We find that the S\&C performance are decoupled when the blocklength $N$ tends to infinity.

Furthermore, it can be seen that the asymptotic expressions of the achievability and converse bound depend on the range of the channel gain, i.e., $|h|_\mathrm{L}$ and $|h|_\mathrm{U}$. When we have more accurate prior knowledge of the channel gain under the assistance of sensing, the interval $[|h|_\mathrm{L},|h|_\mathrm{U}]$ approaches to the true channel gain $|h|$, which also implies that the achievability and converse bound coincides, i.e.,
\begin{equation}
 \lim_{N\to \infty}R^\mathrm{L}(N,\epsilon,D) =  \lim_{N\to \infty}R^\mathrm{U}(N,\epsilon,D) = C
\end{equation}
where $C = \log_2(1+\rho|h|^2/\sigma^2) = \log_2(1+\mathrm{SNR})$ is the Shannon channel capacity.

\begin{figure}[t]
    \centering
    \includegraphics[width = 9cm,height = 7cm]{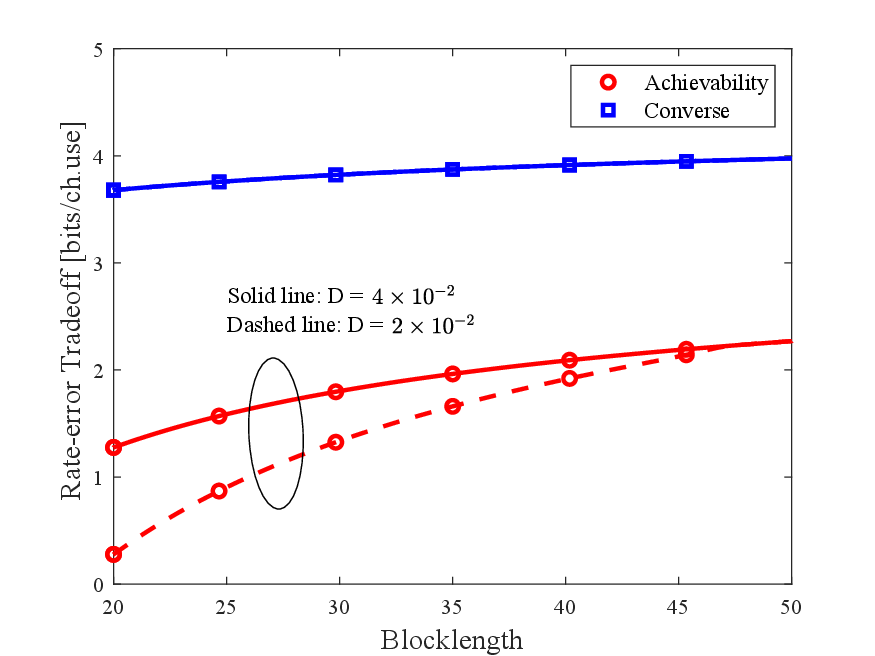}
    \caption{ The achievability and converse bounds for the rate-error tradeoff with varying code blocklength}
    \label{Fig_rate_error}
\end{figure}

\section{Simulation Results}
In this section, we perform some simulation experiments to consolidate our theoretical bounds and calculate the rate-error region numerically. 

First, we verify the effectiveness of the achievability and converse bounds derived for the rate-error tradeoff $R^\star(N,\epsilon,D)$. Consider the signal model given by \eqref{Channel}, where the per-codeword power budget and the noise variance are set as $\rho = 10$ and $\sigma = 1$, respectively. The channel gain is assumed to belong to the set $|h|\in[1,1.5]$. The achievability and converse bounds of the rate-error tradeoff $R^\star(N,\epsilon,D)$ with varying blocklength $N$ is shown in Fig. \ref{Fig_rate_error}. The probability of decoding error is set to be $\epsilon = 10^{-3}$, while the sensing performance is set to be $D = 4\times10^{-2}\ \mathrm{and}\ 2\times10^{-2}$, respectively.

According to Fig. \ref{Fig_rate_error}, we find that the converse bounds always outperform the achievability bounds, which is invariant with the sensing performance $D$ since the loss term almost vanishes according to our theoretical analysis. As for the achievability bounds, the rate-error tradeoff increases as $D$ increases since the sensing constraint is relaxed. When $N$ is large enough, the S\&C performance is decoupled, which implies that the two achievability bounds converge to the same value $\tilde{R}^\mathrm{L}_\mathrm{com}(N,\epsilon)$.

\begin{figure}[t]
    \centering
    \includegraphics[width = 9cm,height = 7cm]{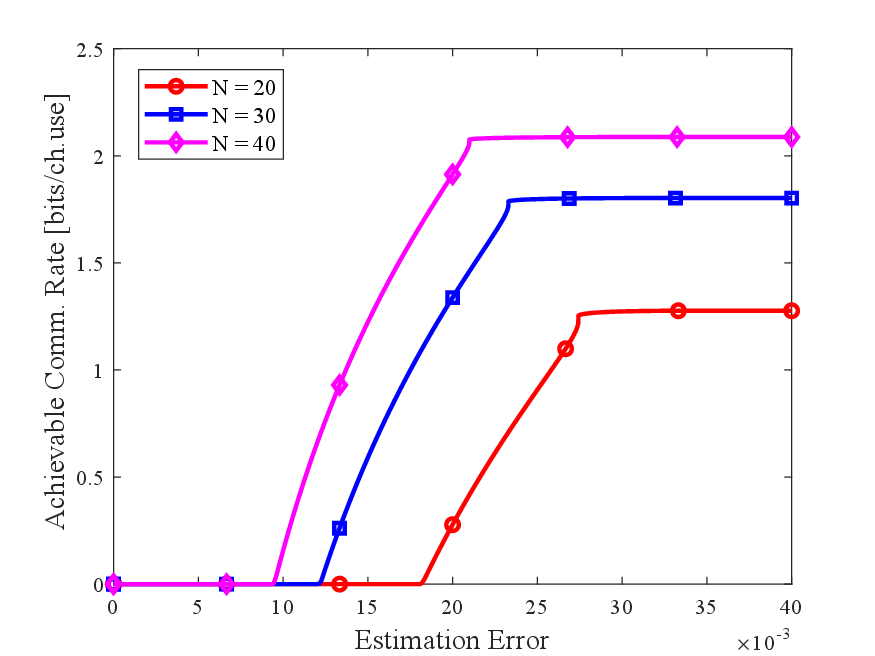}
    \caption{ The achievable rate-error region with the varying blocklength}
    \label{Fig_region}
\end{figure}

Then we calculate the rate-error region $\mathcal{F}(N,\epsilon)$ numerically, which are based on the achievability bound since it can provide a more accurate characterization of the performance tradeoff between S\&C than the converse bound. The system parameters are set the same as above. The achievable rate-error region is shown in Fig. \ref{Fig_region} with the blocklength set to be $N = 20,30,40$, respectively.

According to Fig. \ref{Fig_region}, we find that as $D$ increases, the achievable rate $R$ remains to be zero at first. When $D$ exceeds the threshold bigger than $\sigma^2/N\rho$, the achievable rate $R$ starts to increase since the achievability bound requires the feasible codeword set $\mathcal{W}$ large enough to carry information. As $D$ moves close to $D_\mathrm{m}$, the coefficient $\gamma_\mathrm{L}$ approaches $1/2$ according to \eqref{Achievability-rate-loss}. However, it switches from $1/2$ to $1$ when $D$ moves past $D_\mathrm{m}$, which leads to the $1/N$ sharp increase in the rate-error tradeoff shown in Fig. \ref{Fig_region}. Then the boundary is invariant of $D$, indicating that the S\&C performance is decoupled.

Furthermore, the area of the rate-error region increases with the blocklength $N$, since the performance tradeoff between S\&C vanishes in the large blocklength regime. As the blocklength $N$ tends to infinity, the boundary of the achievable rate-error region approaches the horizontal line with height $\log_2(1+N\rho|h|_\mathrm{L}^2/\sigma^2) $.

\section{Conclusion}
This paper provides a characterization of the performance tradeoff between S\&C in a SISO ISAC system with finite blocklength where the rate-error tradeoff is introduced as the performance metric. In particular, we derive the achievability and converse bounds for the rate-error tradeoff, after which the asymptotic analysis is performed to show that the performance tradeoff vanishes as the blocklength tends to infinity. Finally, our theoretical results are verified by the numerical experiments. Future work will focus on obtaining tighter bounds for the rate-error tradeoff as well as the extension to MIMO ISAC systems. The contributions of this paper give insights to the understanding of the fundamental tradeoff and the future system design in ISAC.



\bibliographystyle{IEEEtran}
\bibliography{IEEEabrv,StringDefinitions,SGroupDefinition,SGroup}

\begin{thebibliography}{10}
\providecommand{\url}[1]{#1}
\csname url@samestyle\endcsname
\providecommand{\newblock}{\relax}
\providecommand{\bibinfo}[2]{#2}
\providecommand{\BIBentrySTDinterwordspacing}{\spaceskip=0pt\relax}
\providecommand{\BIBentryALTinterwordstretchfactor}{4}
\providecommand{\BIBentryALTinterwordspacing}{\spaceskip=\fontdimen2\font plus
\BIBentryALTinterwordstretchfactor\fontdimen3\font minus
  \fontdimen4\font\relax}
\providecommand{\BIBforeignlanguage}[2]{{%
\expandafter\ifx\csname l@#1\endcsname\relax
\typeout{** WARNING: IEEEtran.bst: No hyphenation pattern has been}%
\typeout{** loaded for the language `#1'. Using the pattern for}%
\typeout{** the default language instead.}%
\else
\language=\csname l@#1\endcsname
\fi
#2}}
\providecommand{\BIBdecl}{\relax}
\BIBdecl

\bibitem{HugKawSim:J15}
H.~Griffiths, L.~Cohen, S.~Watts, E.~Mokole, C.~Baker, M.~Wicks, and S.~Blunt,
  ``\textnormal{Radar Spectrum Engineering and Management: Technical and
  Regulatory Issues},'' \emph{Proc. {IEEE}}, vol. 103, no.~1, pp. 85--102, Jan.
  2015.

\bibitem{LiuCuiMas:J22}
F.~Liu, Y.~Cui, C.~Masouros, J.~Xu, T.~X. Han, Y.~C. Eldar, and S.~Buzzi,
  ``\textnormal{Integrated Sensing and Communications: Toward Dual-Functional
  Wireless Networks for 6G and Beyond},'' \emph{{IEEE} J. Sel. Areas Commun.},
  vol.~40, no.~6, pp. 1728 -- 1767, Jun. 2022.

\bibitem{StuWie:J11}
C.~Sturm and W.~Wiesbeck, ``\textnormal{Waveform Design and Signal Processing
  Aspects for Fusion of Wireless Communications and Radar Sensing},''
  \emph{Proc. {IEEE}}, vol.~99, no.~7, pp. 1236 -- 1259, Jul. 2011.

\bibitem{ZhaRahWu:J21}
J.~A. Zhang, M.~L. Rahman, K.~Wu, X.~Huang, Y.~J. Guo, S.~Chen, and J.~Yuan,
  ``\textnormal{Enabling Joint Communication and Radar Sensing in Mobile
  Networks—A Survey},'' \emph{{IEEE} Commun. Surveys Tuts.}, vol.~24, no.~1,
  pp. 306 -- 345, Oct. 2021.

\bibitem{AhmKobWig:J22}
M.~Ahmadipour, M.~Kobayashi, M.~Wigger, and G.~Caire, ``\textnormal{An
  Information-Theoretic Approach to Joint Sensing and Communication},''
  \emph{{IEEE} Trans. Inf. Theory}, 2022.

\bibitem{ChaMoeHim:J17}
B.~K. Chalise, M.~G. Amin, and B.~Himed, ``\textnormal{Performance Tradeoff in
  a Unified Passive Radar and Communications System},'' \emph{{IEEE} Signal
  Process. Lett.}, vol.~24, no.~9, pp. 1275 -- 1279, Sep. 2017.

\bibitem{XioLiuCui:J22}
Y.~Xiong, F.~Liu, Y.~Cui, W.~Yuan, T.~X. Han, and G.~Caire, ``\textnormal{On
  the Fundamental Tradeoff of Integrated Sensing and Communications Under
  Gaussian Channels},'' \emph{arXiv preprint arXiv:2204.06938}, 2022.

\bibitem{PolPooVer:J10}
Y.~Polyanskiy, H.~V. Poor, and S.~Verdu, ``\textnormal{Channel Coding Rate in
  the Finite Blocklength Regime},'' \emph{{IEEE} Trans. Inf. Theory}, vol.~56,
  no.~5, pp. 2307 -- 2359, May 2010.

\bibitem{VanBel:B68}
H.~L.~V. Trees and K.~L. Bell, \emph{\textnormal{Detection Estimation and
  Modulation Theory, Part 1}}.\hskip 1em plus 0.5em minus 0.4em\relax New York,
  NY, USA: Wiley, 1968.

\bibitem{ZhaLiuMas:J21}
J.~A. Zhang, F.~Liu, C.~Masouros, R.~W. Heath, Z.~Feng, L.~Zheng, and
  A.~Petropulu, ``\textnormal{An Overview of Signal Processing Techniques for
  Joint Communication and Radar Sensing},'' \emph{{IEEE} J. Sel. Topics Signal
  Process.}, vol.~15, no.~6, pp. 1295 -- 1315, Nov. 2021.

\bibitem{YanDurKoc:C13}
W.~Yang, G.~Durisi, T.~Koch, and Y.~Polyanskiy, ``\textnormal{Quasi-static SIMO
  fading channels at finite blocklength},'' in \emph{Proc. IEEE Int. Symp. on
  Inf. Theory}, Istanbul, Turkey, Jul. 2013.

\bibitem{YanDurKoc:J10}
------, ``\textnormal{Quasi-Static Multiple-Antenna Fading Channels at Finite
  Blocklength},'' \emph{{IEEE} Trans. Inf. Theory}, vol.~60, no.~7, pp. 4232 --
  4265, Jul. 2014.

\bibitem{Sha:J59}
C.~E. Shannon, ``\textnormal{Probability of error for optimal codes in a
  Gaussian channel},'' \emph{Bell Syst. Tech. J.}, vol.~38, no.~3, pp.
  611--656, May 1959.

\bibitem{FozMcLSch:J03}
M.~Fozunbal, S.~McLaughlin, and R.~Schafer, ``\textnormal{On performance limits
  of space-time codes: a sphere-packing bound approach},'' \emph{{IEEE} Trans.
  Inf. Theory}, vol.~49, no.~10, pp. 2681 -- 2687, Oct. 2003.

\end{thebibliography}



%

\end{document}